# Microdischarge dynamics of volume DBD under the natural convection flow

Y.A. Ussenov[1,2], A.I. Ashirbek[1,2], A.R. Abdirakhmanov[3], M.K. Dosbolayev[3], M.T. Gabdullin[4], T.S. Ramazanov[3]

[1]NNLOT, Al-Farabi KazNU, Almaty, Kazakhstan
[2]Institute of Apllied Sciences and IT, Almaty, Kazakhstan
[3]IETP, Al-Farabi KazNU, Almaty, Kazakhstan
[4]Kazakh British Technical University, Almaty, Kazakhstan

## Abstract

The impact of convective flow due to the temperature gradient between discharge cell electrodes and ambient air was studied for parallel-plate electrode arrangement. It was shown that an increase in temperature during the discharge operation leads to an increase in MD number in the discharge gap. The convective gas flow velocity also increased at a higher thermal gradient, and MD channels follow the gas flow direction. The velocity of convective flow was estimated by the CFD simulation and compared with the mean velocity of MD channels obtained by the particle image velocimetry method.

## Introduction

Non-thermal atmospheric pressure plasmas are widely used in diverse technology fields such as plasma medicine and agriculture, plasma material deposition, surface modification, and additive manufacturing, plasma catalysis, and active flow control (Brandenburg, 2017; Kogelschatz, 2004). The dielectric barrier discharge (DBD) is the source of nonthermal atmospheric pressure plasmas which could operate at open-air room conditions without any vacuum chambers and equipment. Compared to other conventional atmospheric pressure plasmas sources, both (or at least one) of the insulating barrier electrodes of the DBD covers the metal electrodes surface, thus prevents the glow to arc transition by limiting the current. The DBD at room conditions, operating in the air as the plasma forming gas, usually exists in the form of randomly distributed thin discharge channels. These thin (~100-200 μm in diameter) discharge channels are called "microdischarges" (MD) and such type of the DBD can be referred as "filamentary". MD channels in the filamentary DBD are distributed in the space between electrodes and have a lifetime of few tens of nanoseconds. Due to the insulating dielectrics and capacitive nature of discharge, the DBD is generated with periodic sinusoidal or pulsed high voltage signals.

MD can be generated when some threshold local electric field value is reached in the discharge gap, thus each of them can be considered as independent discharge channels. However, usually the dynamics of MD channels in DBD have complex behavior and depends on the collective interaction between them. Volume air DBD with symmetric discharge electrodes generated by sinusoidal high voltage (HV) signal contains stochastically distributed MD channels. MD channels show advanced collective phenomena and form various self-organized patterns and structures at some specific conditions, e.g. lower operating pressure (100 - 300 Torr), noble gas dilution, cryogenic temperature, temperature gradient on the electrodes (Dong et al., 2008). Collective interaction and dynamics of MD channels are very important not only for static volume DBD in the air, but also for DBD under the action of external factors such as gas flow (Wang et al., 2020). Furthermore, additional gas flow could be considered as the simple and cheapest route to the uniform diffuse air DBD, which is extremely desirable for technological applications(Trelles, 2016). Therefore, a thorough attention was made in the last decades by different groups into the investigation of DBD under the airflow to shine a light on the collective dynamics of MD channels and overall changes in the operating regimes of discharges.

Impact of induced airflow (up to 110 m/s) on spatial and temporal distributions of MD filaments and breakdown characteristics of discharge were studied in parallel-plate volumetric AC (15-25 kHz) excited DBD (Fan et al., 2016). The MD filaments follow the direction of the airflow, and their average velocity was lower than the corresponding airflow velocity. Sequential increasing in gas flow leads to changes in spatial distribution of MD channels and formation of at least four discharge phases. The complex change in spatial structure and collective behavior of MD channels at different gas flow regimes is explained by the impact of forced convection on the discharge remnants and redistribution of charged, excited particles. This redistribution results in different force balance between MD remnants. Along with changes in structure of discharge filaments, the electrical, optical, and thermal properties are also subject to changes. In particular, the airflow leads to the decrease in breakdown voltage and gas temperature, while the intensity of MD filaments increases. A study of a nanosecond barrier discharge under the influence of forced gas flow also shows a change in the discharge modes (Fan et al., 2019). A gradual increase in the flow rate leads to a transition of the discharge regime from filamentary to diffuse. After reaching a certain critical high velocity, the discharge transitions back to filamentary, but with a noticeably small diameter of the MD channels.

The gas flow convection in discharge gap determines the spatial and temporal distribution of charged and excited particles, which in turn are responsible for pre-ionization in the MD channels in DBD. Pre-ionization in the discharge gap is a key factor in further evolution of individual microdischarges and formation of collective phenomena as self-organized patterns, spatial "memory effect" of microdischarge channels (Höft et al., 2016). Even if there is no external gas flow, there is always room for particle transport due to intense heat transfer and natural convection, since microdischarges in the dielectric gap are a local source of heat. Although the DBD is a transient thermal nonequilibrium discharge, Joule heating in a couple of chemical dissipation of energy in the discharge volume, and dissipation of part of the applied power as heat on dielectrics leads to active heating of the walls of the discharge cell. While the volume DBD behavior in an externally stimulated gas flow has been extensively studied, there are insufficient data on the impact of the natural convection gas flow on the collective behavior of MD channels and structure of the filamentary barrier discharge in air (Caquineau et al., 2009).

In this contribution we investigate the impact of the self- induced natural convection flow on the collective dynamics MD channels of filamentary DBD in air. The MD behavior analyzed and compared for both vertical and horizontal arrangement of symmetrical DBD electrodes. The particle image velocimetry (PIV) and CFD software were utilized for MD trajectory, velocity analysis and for simulation of buoyancy forces induced gas flow respectively. Electrical properties were also studied for different temperature of DBD electrode walls and corresponding natural convection induced flow velocities.

**Methods**

*Experimental part*

The schematic view of the experimental setup is shown in Figure 1A. The main part of setup is DBD discharge cell, which consists of two parallel conductive electrodes covered with dielectric quartz plates with size 60x60 mm. High voltage electrode is a rectangular aluminum metal foil with size 20 x 60 mm, while transparent and conductive indium tin oxide (ITO) covered glass plate with size 20x60 mm used as grounded electrode. The edges of high voltage aluminum electrode were covered by high thermal resistive silicon past to avoid any parasitic discharges. The sheet resistance of ITO is 8-10 Ohm/cm$^2$, the dielectric permittivity for quartz is ε = 3.0 and for glass is ε = 2.5. The discharge gap, defined as the distance between two quartz plates, is *d = 3 mm and* the plates fixed by two 3D printed plastic (ABC) spacers for the all experiments. The high voltage power source composed of SDX waveform generator (Siglent, Hong-Kong), Pro -Lite 5.0 power amplifier (Crest Audio, USA) and high voltage transformer 2500S (Amazing LTD, USA). The 30 kHz AC sinusoidal signal from waveform generator amplified by audio amplifier and enhanced to high voltage signal by transformer and applied to electrodes of volume DBD. The typical applied voltage and current are 26 kVpp AC and 100 mA, respectively. The I-V curve monitored by Wave Jet 354A oscilloscope (LeCroy, France). High voltage detected by Tektronix P6015 -1/1000 (Tektronix, USA) probe, while current detected by low voltage probe through low-inductance $R$=51 Ohm shunting resistor. The IR camera is used to detect the change of DBD quartz dielectric electrode temperature. The IR camera placed always in perpendicular to ITO glass electrode and measures the outer side of quartz plate due to the low emissivity and reflective behavior of ITO surface. The field of view of IR camera contains the active part of the discharge cell where the MD channels exist, and the size is 20x60 mm. As the reference temperature the temperature of the quartz plate at the ITO glass border was taken. Additional K type thermocouple utilized for reference temperature measurement and fixed by thermal grease on the surface of quartz plates in close vicinity of ITO glass electrode. The high-speed camera Phantom VEO 710-S (Phantom, USA) applied to detect the video frames with 500 fps and exposure time 2 ms. The long exposure (up to 41 ms) images were obtained by Nikon CX41 photo camera. The experiments conducted at open air room conditions with 300 K temperature, 45 % absolute humidity and ~ 710 Torr base pressure.

After the discharge generation, the temperature of the quartz dielectric surface close to the ITO glass grounded electrode was continuously measured with an IR camera. The first video images of MD with a length of 1 s were recorded immediately with a high-speed video camera at 500 frames per second, while subsequent video frames (also with a length of 1 s) were recorded at intervals of 30 seconds. For each of the video frames, the corresponding I-V curves was recorded using a digital oscilloscope and stored to PC. The experiments were performed for both horizontal and vertical arrangements of the discharge cell. For horizontal arrangement the gravity force and natural convection flow direction was parallel to MD propagation channel, while for vertical arrangement the MD filaments was perpendicular to gravity and convective flow direction (Figure 1B). Before each experimental run the discharge cell cooled down to room temperature about 30 min.

*Natural convection flow simulation.*

The natural convection flow in the vertically oriented discharge gap, induced by heat transfer from MD plasma channels to discharge walls, simulated by Comsol Multyphics Package. The heat and mass transfer modules with weakly incompressible and laminar flow was applied. The system considered in 2D arrangement, and the flow only in active discharge cell part with size 20 mm x 3 mm were analyzed. The temperature gradient considered as difference between quartz plates temperature and ambient air with constant temperature. Therefore, the open boundary conditions applied, and the overall computational mesh size was equal to $10^3$ cells. The obtained distribution of natural convection velocity values was compared with MD channels velocities.

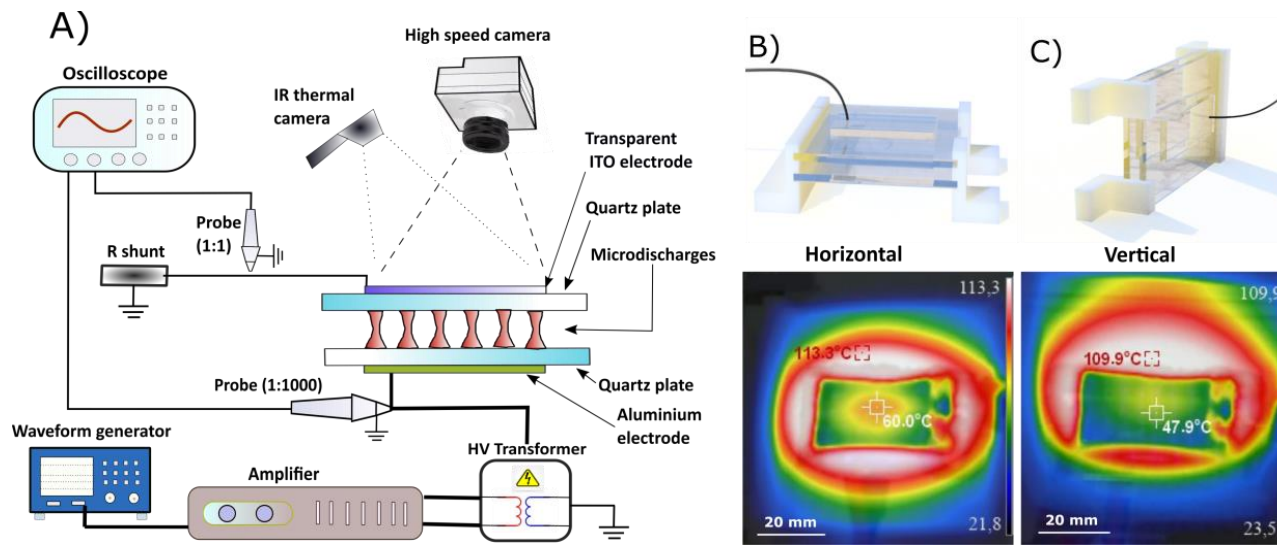

Figure 1. General scheme of experimental setup (A) and arrangement of DBD discharge cell at horizontal (B) and vertical (c) position.

**Results and discussion**

The heating of the discharge cell wall continuously monitored by the IR thermal camera throughout the one single experimental run. As an example, the typical samples of IR camera images for horizontal and vertically arranged electrodes at 270 s were also shown in Figure 1B and Figure 1C. Due to the high reflectivity of ITO covered top glass electrode, as a reference, the quartz barrier electrode temperature was taken. Moreover, as an internal electrode, the heating dynamics of quartz electrode has the direct impact on heat transfer to the discharge gap volume. The shift of the heating spot to the upward direction in vertically arranged discharge cell confirms the existence of buoyancy flow due to free convection of air inside the discharge gap.

The change in the temperature of dielectric quartz electrode during the 300 s operation of the discharge for both vertical and horizontal arrangement is shown in Figure 2. The presented data corresponds to the highest temperature point of quartz dielectric plates in the vicinity of grounded ITO covered glass electrode and obtained from the IR camera images. The difference between this maximum temperature and ambient air temperature (~22 $^0$C) defined as $\Delta T_{max}$. After the plasma breakdown, the dielectric discharge wall temperature increases gradually from 295 K to 113,3 K for horizontal and up to 109,9 K for vertical adjusted discharge cells. The slightly slow rise of temperature for vertical discharge cell arrangement can be explained by more effective cooling of discharge gap by natural convection.

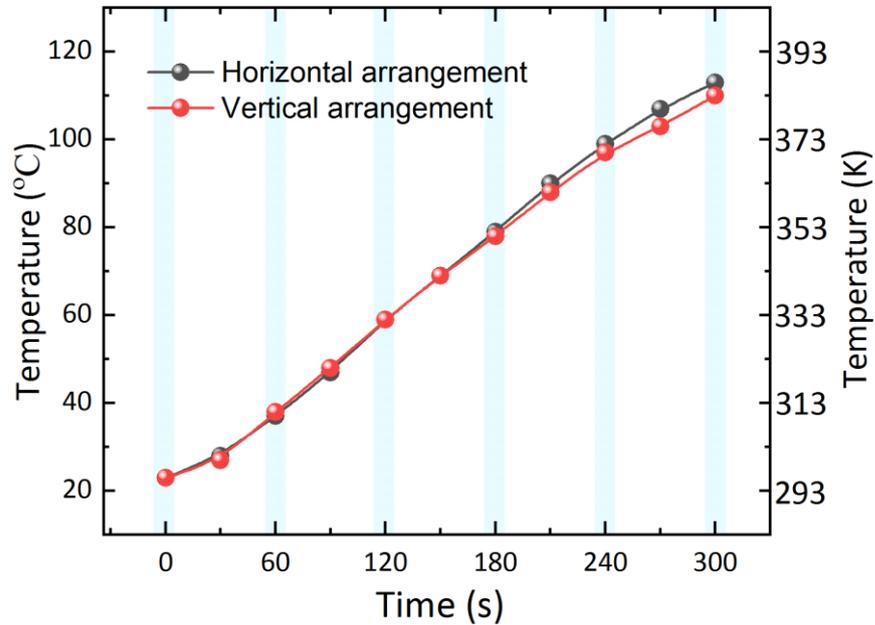

Figure 2. The change in the temperature of dielectric quartz electrode during the 300 s operation of the discharge for both vertical and horizontal arrangement

The obtained top view photo images of discharge structure and MD channels evolution are demonstrated in Figure 3. The visual images taken by Nikon photo camera at 24 frames per second (fps) and 41 ms exposure time. In images, the MD channels appears as the light spots with average diameter of 500-1000 μm on the inner surface of ITO glass covered quartz plates. After 180 s of operation the bright halo like cloud arise around each MD channel spots, what is the consequence of intensive surface ionization waves.  As show the photos, an increase in temperature of cell wall during the operation of the discharge leads to an increase in the number of MD over time. At horizontal position of the discharge cell, the motion of MD channels is random and the increase in MD number leads to shorten travelling distance and the distance between each MD channel. As the discharge cell acts as a trap for MD channels, at higher discharge cell temperature, the MD channels only oscillate around their initial position.

For the vertical arrangement of the discharge cell MD channels show more complex behavior. At the first 60 s of discharge operation, the MD channels collective motion have the similar behavior as for the horizontal arrangement. The MD channels slowly migrate to the random direction, with oscillation of channels around the own central axis. However, rise of the discharge cell wall temperature and the thermal gradient up to $\Delta T_{max}= 56.2$ K results in directed motion of MD channels by following the natural convection flow direction. As a result, at the $\Delta T_{max}= 56.2$ K (Figure 3(E)) the formation of bright line shape trajectory started. The accumulation of MD spot position during the 41ms exposure time leads to formation of such line shape path of trajectory. Further increase in thermal gradient up to $\Delta T_{max}= 87.9$ K (300 s) leads in increase of this effect: the MD channels direction is more obvious, the velocity is faster and the length of line shape spots are longer (Figure 3(F)).

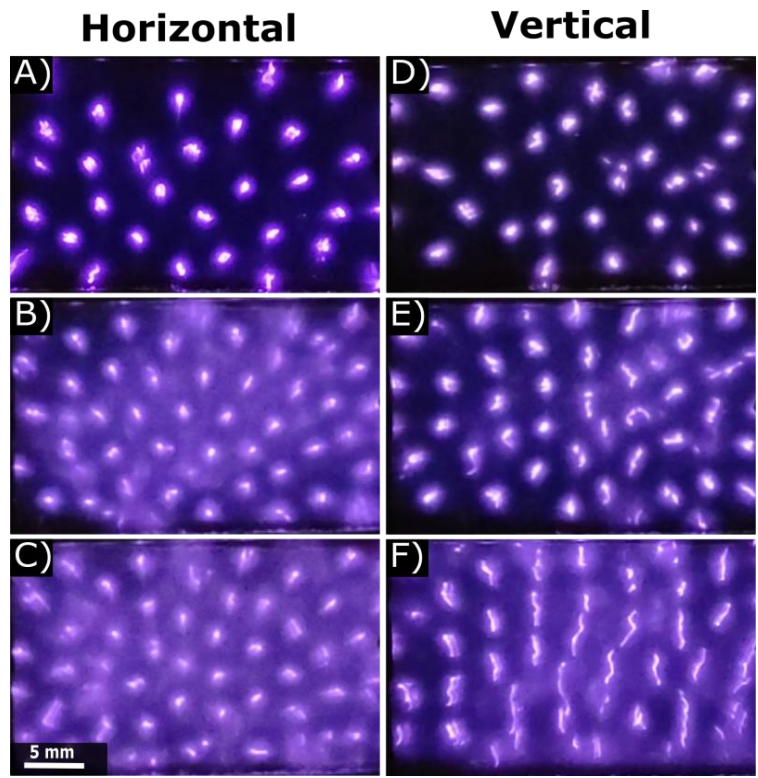

Figure 3. The top view photo of MD dynamics behavior at different discharge operation time and temperature gradient for horizontal: (A) 60 s, $\Delta T_{max}=16.0$ K, (B) 180 s, $\Delta T_{max}=57.2$ K, (C) 300 s, $\Delta T_{max}=91.3$ K, and vertical: (D) 60 s, $\Delta T_{max}=16.0$ K, (E) 180 s, $\Delta T_{max}=56.2$ K, (F) 300 s, $\Delta T_{max}=87.9$ K position of the discharge cell. The camera frame rate is 24 fps, 41 ms exposure time, the discharge gap and frequency are 3 mm and 30 KHz, respectively.

More detailed analysis of the collective dynamics of MD channels was carried out by *particle image velocimetry* method. The motion of MD channels analyzed from 500 images accumulated during the 1s at 60 s, 120s, 180s, 240 s and 300 s time stamps of discharge operation. The exposure time of each frame and time interval between them are 2 ms and 0.1 ms respectively. The MD velocity distribution maps and the motion direction for different $\Delta T_{max}$ corresponding to the 60 s, 180 s and 300 s are presented in Figure 4. Figure 5 provides the velocity distribution histograms. For the horizontal adjusted discharge cell, the motion direction of MD channels is not obvious and have the random nature, which can be seen by diverse vector arrows indicating different direction. The velocity of MD channels $V_{MD}$ increases slightly due to the change in discharge cell temperature and $\Delta T_{max}$ over time. Indeed, the $V_{MD\ distribution}$ histograms for horizontal discharge cell in Figure 5 confirm that the mean value of $<V_{MD}>$ grow from ~ 0.025 m/s up to ~ 0.038 m/s, for the $\Delta T_{max} = 57.2$ K and $\Delta T_{max} = 91.3$ K respectively.

The collective motion behavior of MD channels is completely different for the vertical arrangement of the discharge cell. As the $\Delta T_{max}$ increases from 16.0 K to 56.2 K, almost all the MD filaments move towards the top edge of the cell, which is clearly indicated by the velocity vectors (Figure 5 (E)). Since the buoyancy forces due to the convective gas flow directed controversial to the gravity, the MD channels start to follow it and migrate to the top edge of the discharge gap. After $\Delta T_{max}$ reaches the 87.9 K wthin the 300 s of operation, the MD filaments collective motion direction more pronounced, and MD velocity sufficiently increased (Figure 5 (F)). As show the velocity surface map and speed vectors, the MD channels on the central region of the discharge cell have strong upward direction and higher velocity, while the motion of MD on the left and right sides shifted to the edges and migrate with lower velocity.

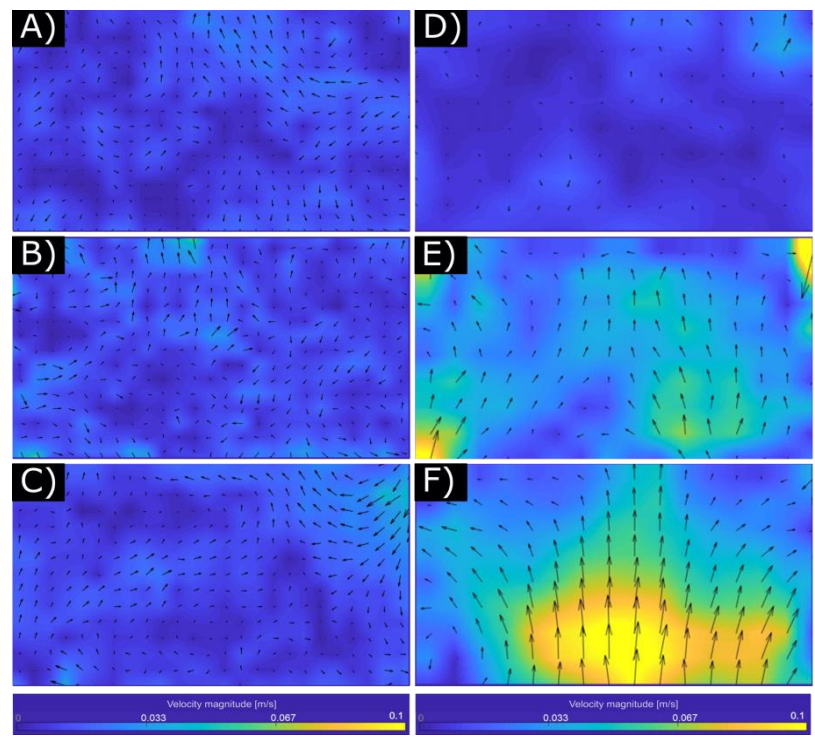

Figure 4. The MD filaments PIV velocity surface distribution map and collective motion

direction vectors obtained by PIV analysis for horizontal: (A) 60 s, $\Delta T_{max}$=16.0 K, (B) 180 s, $\Delta T_{max}$=57.2 K, (C) 300 s, $\Delta T_{max}$=91.3 K, and vertical: (D) 60 s, $\Delta T_{max}$=16.0 K, (E) 180 s, $\Delta T_{max}$=56.2 K, (F) 300 s, $\Delta T_{max}$=87.9 K adjustment of the discharge cell. The camera frame rate is 500 fps, 2 ms exposure time, the discharge gap and frequency are 3 mm and 30 KHz, respectively.

The MD velocity distribution obtained from PIV analysis (Figure 5) for vertical discharge cell shows that for 60 s, $\Delta T_{max}$ = 16.0 K the average $<V_{MD}>$ = 0.015 m/s, which is comparable for horizontal case. Further increase of $\Delta T_{max}$ up to 56.2 K after 180 s has been accompanied by an increase in the $<V_{MD}>$ up to 0.04 m/s. At the same time the maximum $V_{MD}$ reaches ~ 0.115 m/s, indicating the broad width of corresponding Gaussian fit of histogram. The MD mean velocity grow up to $<V_{MD}>$ = 0.05 m/s for the $\Delta T_{max}$ = 87.9 K, demonstrating high impact of well directed convective flow on MD filaments collective motion (Figure 5). While the highest frequency detected for filaments with $V_{MD}$ =0.042 m/s, the velocity range lies in the 0.008 m/s - 0.012 m/s interval.

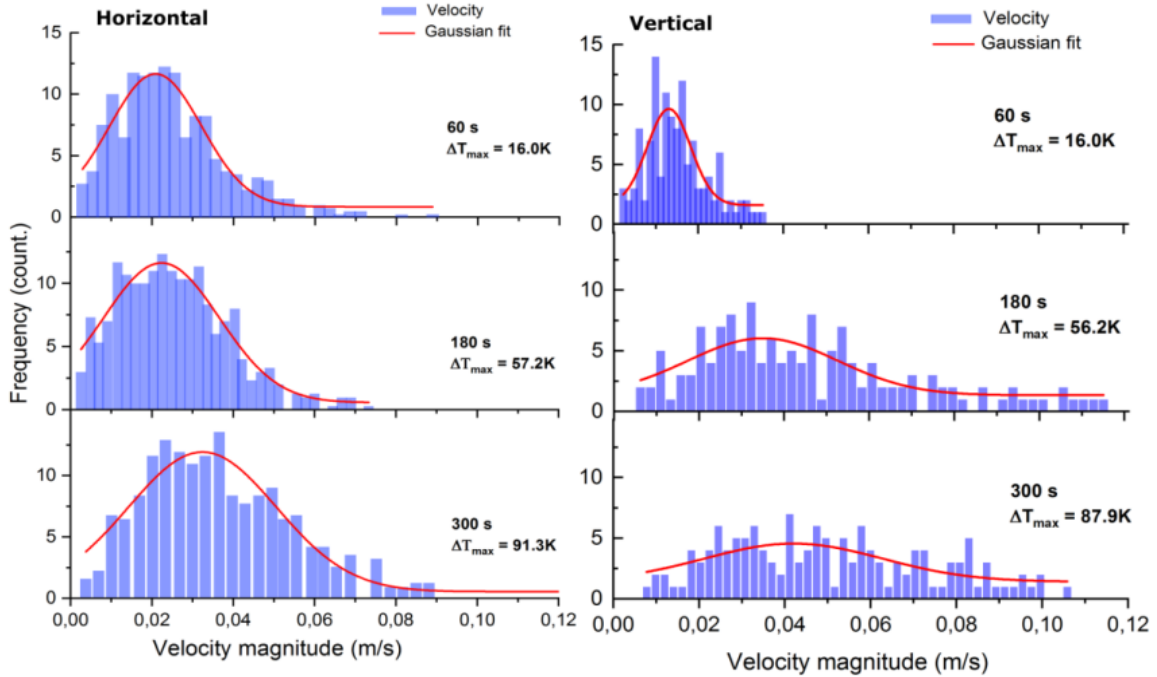

Figure 5. The MD velocity distribution histogram obtained by the PIV analysis for horizontal and vertical arrangement at 60 s, 180 s, and 300 s discharge operation.

The time resolved I-V curve and it's change during the operation of the DBD were also examined for different $\Delta T_{max}$. Figure 6 demonstrates the change in I-V curve at different time stamps for both horizontal and vertical electrode arrangements. It is well known that in DBD, each pulse of current oscillogram corresponds to the formation single MD filament. However, the separate filaments could appear at the same time and in that case the corresponding current pulse show the highest peaks which is the sum of currents transferred by single MD. Therefore, number of detectable current pulses by oscilloscope usually less than number of MD channels captured by long exposure time camera. As noted above, the longer discharge operation time results in intensive heating of the wall and increasing amount of the MD channels. As expected, higher amount of MD results in more peaks in corresponding current pulses (Figure 6). Thus, the average number of detected current pulses per voltage half cycle change from 6 to 11 for horizontal, and from 5 to 10 for vertical arrangement within the 300 s of discharge operation time. The increase in discharge temperature also leads in change the discharge sustaining voltage Vs and average discharge power *P*. Figure 7 provides the evolution of *Vs*, average number of MD and average distance between separate MD channels for both horizontal and vertical arrangements. Average MD amount and interfilament distance were obtained after analyzing 500 discharge frames and each frame has 1/500 s exposure time. Hence, one single image corresponds to accumulation of 120 discharge cycles (voltage half periods). With rise in the $\Delta T_{max}$ the $V_s$ gradually decrease from 12.2 *kV* to 10.75 *kV* within the 300 s discharge operation time. At the same time, the mean number MD of channels grow from ~31 to ~ 42 for both vertical and horizontal arrangement. For the vertical discharge cell, the mean distance between MD channels significantly drops from ~ 4 mm at 60 s to ~ 2.75 mm at 120 s. Further discharge operation and grow in $\Delta T_{max}$ results in gradually decrease down to ~ 2.6 mm at 300 s. In case of horizontally arranged discharge cell the distance first decrease drastically from ~ 4.7 mm down to ~ 3.0 mm at 120 s and gradually drop to ~ 2.7 mm at 300 s of discharge operation time.

The COMSOL simulation was made for vertical discharge cell arrangement to estimate the natural convection flow velocity inside the discharge gap. As mentioned earlier, a static 2D geometry with isothermal discharge walls and open boundary conditions was considered. As the initial wall temperature, the IR camera measured maximum temperature taken into account. As the active surface for simulation, only the part of quartz dielectric plates covered with high voltage metal and grounded ITO glass were chosen. Moreover, the simulation results visualized from the side view (cross section) and clearly demonstrate the air flow velocity distribution inside the discharge gap between two discharge dielectric walls. The existence of MD plasma channels and the heat transfer from each MD are not considered in the COMSOL simulation. The effect of the plasma by hydrodynamic forces was not considered because the airflow is transversal to the electric field. Also, we assume that the heat, generated by separate MD plasma channels, transferred to quartz dielectric walls and only then, after excessive heating these walls,

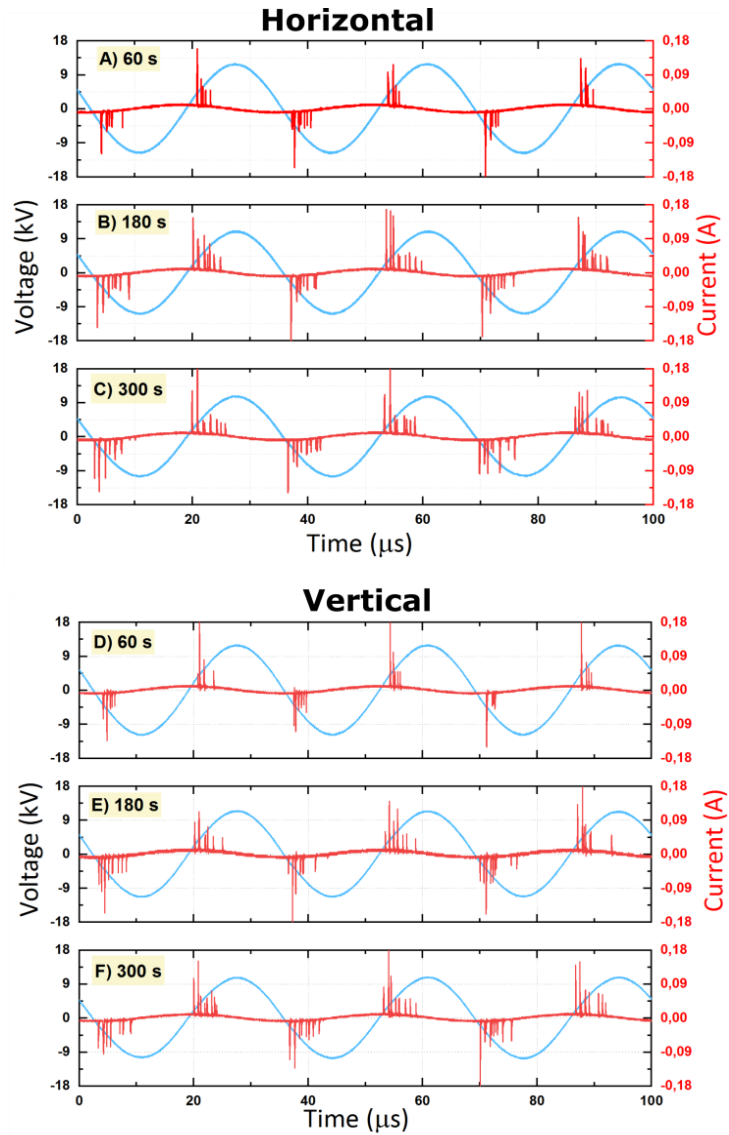

Figure 6. The I-V curve of DBD with horizontal and vertical discharge cell at different operation time points.

The average discharge power <P> calculated by the following formula:

$$<P> = \frac{1}{t_2-t_1}\int_{t_1}^{t_2} I(t)U(t)dt,$$

where $U(t)$ and $I(t)$ the voltage and current respectively. The time interval between $t_1$ and $t_2$ cover 26 periods $T$ of high voltage signal. With increasing the discharge operation time and grow of $\Delta T_{max}$ the <P> decrease gradually form 14 W to 10 W for both horizontal and vertical arrangement of discharge cell.

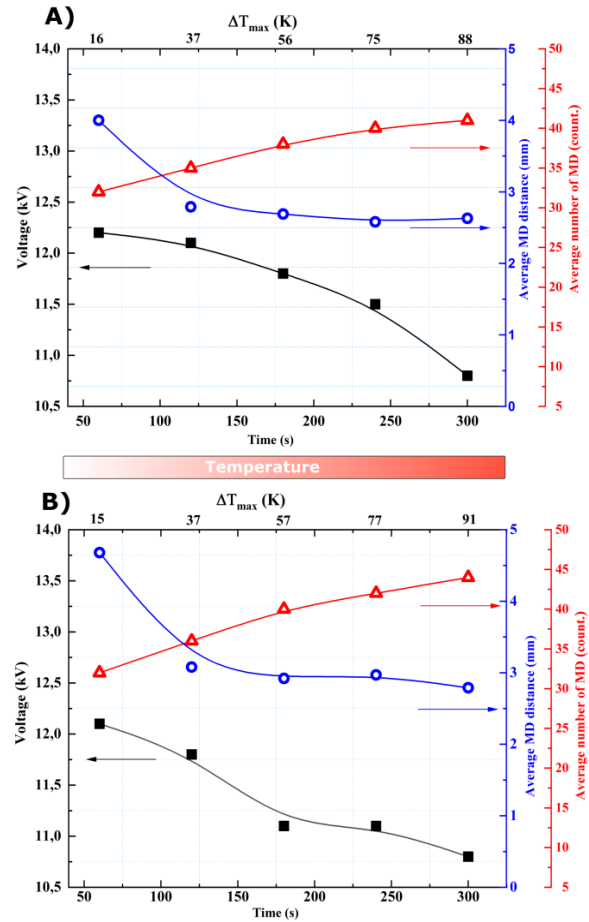

Figure 7. The discharge sustaining voltage $V_s$, mean MD number and average MD channel distance for vertical (A) and horizontal (B) arrangement of discharge cell at different discharge operation time with corresponding $\Delta T_{max}$.

results in generation of stable convective air flow in the discharge gap. Indeed, the Figures 3 and 4 show that the at the beginning of the discharge operation (first 60-120 s), when the discharge wall temperature is not hot enough, there is no stable convective flow which able to change the MD dynamics significantly.

The convective airflow velocity distribution surface map and velocity direction vector arrays in the discharge gap height axis for different $\Delta T_{max}$ at 60 s, 180 s and 300 s discharge time demonstrated in Figure 8A - 8C. Also, the velocity magnitude value distribution between two electrodes provided in Figure 8D. As can been seen from the figures, the horizontal axis velocity distribution inside the vertically adjusted discharge gap has an inverted parabolic shape profile. This is expected due to the high viscosity of flow layers near the discharge walls and common profile for fully developed flow in rectangular duct. The peak value of velocity profiles follow the growing trend of $\Delta T_{max}$ within discharge operation time and rise up to ~ 0.135 m/s for

ΔT$_{max}$ = 87.9 K. For the ΔT$_{max}$=16.0 K and ΔT$_{max}$=56.2 K the corresponding peak velocity values equal to 0.052 m/s and 0.097 m/s respectively. The velocity distribution maps clearly indicate an uniform flow along the electrodes not only inside the discharge gap, but also in the vicinity of the outer surface of the discharge cell.

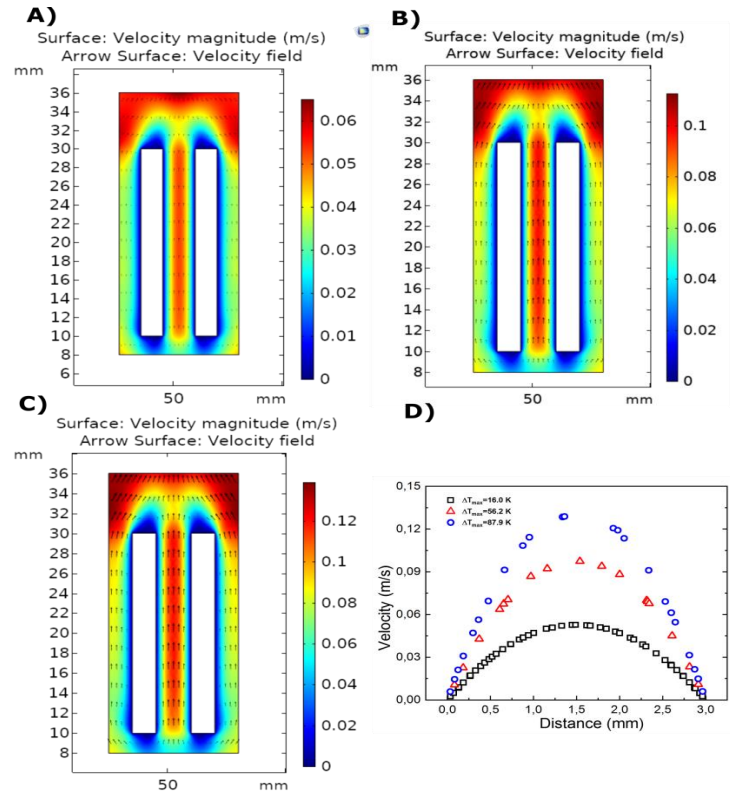

Figure 8. The simulated convective airflow velocity distribution surface map and velocity direction vector arrays in the discharge gap height axis for different ΔT$_{max}$ at 60 s (A), 180 s (B), 300 s (C) discharge time and the corresponding velocity magnitude distribution between two electrodes (D).

The experimentally obtained mean velocity of MD channels and natural convection flow velocity, simulated by Comsol Multiphysics, have been compared in Figure 9. The mean velocity calculated as average velocity of all MD channels obtained by PIV analysis. The convective air flow velocity defined as the peak half-maximum of the velocity profile in the discharge gap. The graph demonstrates that the MD average velocity is lower than convective air flow velocity within the whole 300 s discharge operation time. At 60 s of operation time the mean velocity of MD channels is 0.015 m/s while the simulated airflow velocity is equal to 0.026 m/s. The MD mean velocity slightly grow up to 0.018 m/s for 120 s. Further increase of the discharge operation time up to 180 s and ΔT$_{max}$ = 56.2 K results in jump of mean MD velocity to 0.04 m/s. Note that PIV analysis of MD trajectories show that at this time point of discharge operation the pronounced upward direction of MD channels. The MD mean velocity increase slightly up to 0.5 m/s at 300 s and at the same time the corresponding simulated airflow velocity increase to 0.065 m/s.

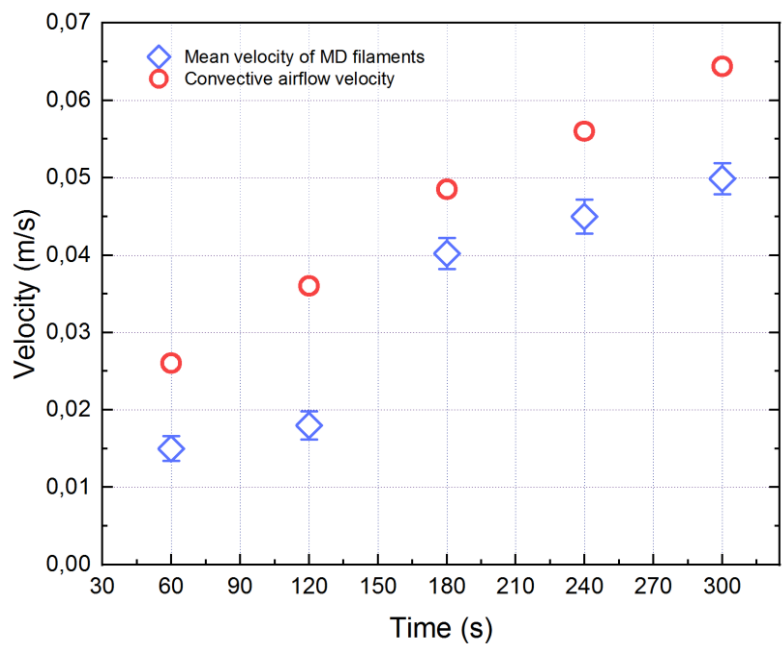

Figure 9. The experimentally obtained mean velocity of MD filaments and simulated natural convection flow velocity.

**Acknowledgements**

This work was supported by the Ministry of Education and Science of the Republic of Kazakhstan, grant number AP09258963